\documentstyle[spie,psfig]{article}
\def\E{{\mathrm E}}

\def\defi{\stackrel{\mathrm def}{=}}
\def\bfr{{\bf r}}
\def\bfv{{\bf v}}
\def\sfr{{\sf r}}
\def\sfv{{\sf v}}
\def\sfG{{\sf G}}
\def\sfS{{\sf S}}
\def\sfa{{\sf a}}

\def\IN{\relax{\rm I\kern-.18em N}}
\def\IR{\relax{\rm I\kern-.18em R}}
\def\calB{{\cal B}}
\def\calC{{\cal C}}
\def\calD{{\cal D}}
\def\calE{{\cal E}}
\def\calF{{\cal F}}
\def\calL{{\cal L}}
\title{Dynamic Models and Nonlinear Filtering\\
of Wave Propagation in Random Fields}
\author{Haiqing Wei\skiplinehalf Gazillion Bits, Inc., San Jose,
CA 95134, USA}  
\begin{document}
\maketitle
\begin{abstract}
In this paper, a general model of wireless channels is established
based on the physics of wave propagation. Then the problems of
inverse scattering and channel prediction are formulated as
nonlinear filtering problems. The solutions to the nonlinear
filtering problems are given in the form of dynamic evolution
equations of the estimated quantities. Finally, examples are
provided to illustrate the practical applications of the proposed
theory.
\end{abstract}

\section{Introduction}
In modern engineering, the problem of modeling wave propagation in
a random field of scattering objects and of the associated signal
processing, has become more and more important, with wide
applications ranging from wireless communications, to radar or
sonar object detection, to medical imaging using microwave or
infrared light. In all of the applications, the first thing is
obviously to understand and model the wave propagation in random
fields. In the object detection and imaging applications, the
signal processing is usually an inverse scattering problem, that
is to find out the configuration and dynamics of the scattering
objects in the random field, based on the data of received signals
and the known transmitted signal. When applied to the field of
wireless communications, for instance, channel estimation and
prediction for dynamic power control or adaptive modulation, the
problem goes one step deeper and becomes filtering and predicting
the dynamics of the channel response, given some observed data of
the past channel response. A powerful model of wave propagation
will be established based upon the physics of wave scattering, by
which the channel response may be computed explicitly as a
functional of the random field of scattering objects. Then, it
becomes natural to approach the problem of estimating and
predicting the channel response in two steps: solving the inverse
scattering problem first, then calculating the channel response
using the solved dynamics of the scattering objects.

Such a natural two-step approach seems to have been overlooked in
the existing literature of communications. In fact, the
traditional approaches try to avoid the inverse scattering
problem. They often start directly from some presumed stochastic
models of the channel response, without looking into the actual
random field of scattering objects at all. In the Rayleigh fading
model that has been widely used since the early days of wireless
communications \cite{Jakes74}, the physical scattering environment
is absent from the consideration, while the received signal is
assumed to be a sum of many differently delayed and weighted
versions of the transmitted signal, where the weighting
coefficients are complex Gaussian and independent, the delay times
modulo $1/f_c$ are also independent and uniformly distributed in
$[0,1/f_c)$, with $f_c$ being the center frequency of the carrier
\cite{Proakis00}. When there is one strong, dominating component,
{\it e.g.} from line-of-sight propagation, the Rayleigh model is
slightly modified into the Ricean fading model. Both models are
based on the assumption of a large number of scattering objects,
which may not always be the case. In case an adaptive mechanism is
employed to control the transmitter power or vary the transmission
bit rate, the channel estimation and prediction circuit has to be
very fast to follow the dynamics of the channel response
\cite{Ziegler92}. The previous channel models have become
inadequate in millimeter wave mobile communications, where the
channel is fast-fading, namely, the bit duration is short
comparing to the dynamics of channel response \cite{Biglieri98}.
In fact, the inadequacy of the old models may be held responsible
for a general opinion among many wireless engineers that in many
fast-fading systems, it might be impossible to estimate the
channel state information (CSI) so to adapt the system accordingly
\cite{Hochwald00}. Such a pessimistic view has even inspired a lot
of work concerning the capacity and coding schemes of future high
speed and fast-fading wireless channels
\cite{Hochwald00,Abou-Faycal01}. A generally used estimation of
the fading speed is based upon the assumption that the signal
fading experienced by two receivers become uncorrelated when the
separation between the receivers is significantly larger than the
carrier wavelength \cite{Salz94}. However, in the Rayleigh or
Ricean model, the signal fading is governed by a Gaussian law of
distribution, which implies the equivalence between uncorrelation
and independence. So even in principle, the signal fading seen in
one region does not contain any information for the signal fading
in another region just a few wavelengths away. This imposes a
serious limitation upon the estimation of CSI. For instance,
consider a wireless link with the carrier frequency around 1.9
GHz, so the wavelength is about 0.158 meter. It takes merely 8
milliseconds for a mobile with speed 20 m/s to travel through one
wavelength. It seems, according to the Rayleigh and Ricean models,
that an estimator for the CSI has to accomplish its task well
within a millisecond, and the obtained CSI can be regarded as
reliable for no longer than 8 milliseconds.

We think that the above limit on the CSI estimation of a
fast-fading channel is over pessimistic. Our point of view is that
for any wireless link in a given physical environment, the channel
response is the result of the superposition of waves propagating
via multiple paths, where each signal path is a geometric route
that starts from the transmitter, possibly hits one or more
scattering objects, and finally reaches the transmitter. The
physical environment that contains the scattering objects is
called a scattering environment. The response of a wireless
channel is actually deterministic, conditioned upon the scattering
environment and the locations of the transmitter and receiver
inside the scattering environment. The randomness of wireless
channels arises only from the variety of scattering environments.
In a wireless channel model based on the physics of wave
propagation in a scattering environment, a natural choice for the
``channel state'' would be the state of the scattering
environment, namely the positions and velocities of the scattering
objects as well as their physical effects on the wave. So the
estimation of CSI boils down to estimating the positions,
velocities, and attenuation coefficients of the scattering
objects. Such channel state should remain largely time-invariant
over a much longer time period. As an example, consider a
scattering environment in which the distance among the scattering
objects, the transmitter, and the receiver is at least several
tens of meters, so that the mobile receiver or transmitter
experiences the same scattering environment even after it travels
over 10 meters. With the 20 m/s mobile speed, it takes 0.5 second
to cover that distance, which is more than 60 wavelengths for a
carrier frequency around 1.9 GHz.

There have been previous works adopting a deterministic model of
signal fading \cite{Eyceoz98,Duel-Hallen00,Hwang98}, but in most
of them the effect of the scattering environment is represented by
a superposition of plane waves in proximity to the receiver, whose
interference pattern is sampled by the receiver. This
representation only works for a frequency-nonselective channel
\cite{Proakis00}, where the frequency bandwidth of the signal is
too narrow to resolve the delay times of different paths. When the
signal bandwidth is sufficiently large to resolve the path delays,
the plane wave representation breaks down, a more advanced model
that captures both the direction of arrivals and the path delays
is necessary. Moreover, the previous works
\cite{Eyceoz98,Duel-Hallen00,Hwang98} estimate the CSI using
various spectral estimation techniques, which do not guarantee the
optimality of the solutions. In this paper, we shall first
establish a complete model of wireless channels based on the
physics of wave propagation. Then the problems of inverse
scattering and channel prediction are formulated as nonlinear
filtering problems. The solution to the nonlinear filtering
problem will be given in the form of a dynamic evolution equation
(the filtering equation) for the estimated quantities, just like
the Kalman filtering equation for the linear filtering problem
\cite{Kailath00}. In most practical applications, the filter
equation may be implemented, or at least be approximated, by a
state-possessing machine driven by an innovation input. In our
particular applications, it happens that both the state of the
scattering environment and the observation process are governed by
linear differential equations. Note that even though the system
equations are linear, the Kalman filter may still be sub-optimal,
unless the initial probability law of the system state is
Gaussian. It is proven mathematically that the nonlinear filter
produces the optimal estimation for the interested quantity. Even
the nonlinear filter may not be easily implemented in practice,
its solution provides a benchmark for evaluating the performance
of other sub-optimal estimators. The theory of nonlinear filtering
has been well established, many textbooks exist, covering its
fundamental principles and applications, especially to the field
of control engineering \cite{Krishnan84,Bensoussan92,Ahmed98},
however it has not made much appearance in the field of
communication engineering. Apart from an attempt to solve the
channel prediction problem for fast-fading wireless channels, the
present paper is intended to introduce the nonlinear filtering
theory as a powerful tool for signal processing and system
optimization in the field of communication engineering.

\section{Random Scattering Fields and Wave Propagation Therein}
\label{waveinrsf} 
\begin{figure}[ht]
\centerline{\psfig{figure=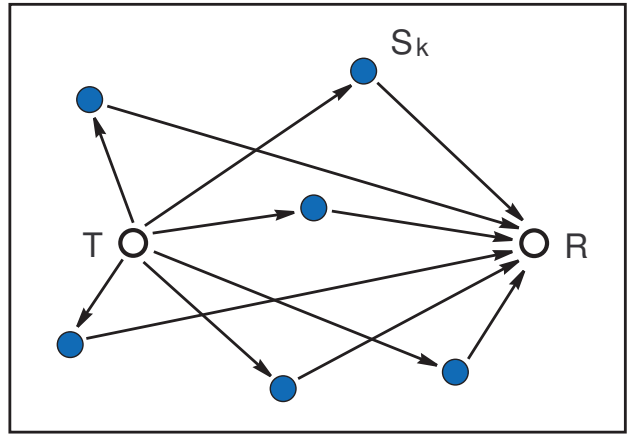,width=11cm}} \caption{A
typical random scattering field, in which the signal wave
propagates from a transmitter to a receiver along various
geometric paths as a result of the interaction with the scattering
objects.} \label{rsf_points}
\end{figure}

Throughout the present paper, a sufficiently
large probability space $(\Omega,\calF,P)$ is assumed in which all
random variables under consideration are defined. In the
wave-involved applications mentioned in the beginning of the
introduction, there is often a random scattering field (RSF)
consisting of scattering objects, due to which the signal wave
from a transmitter to a receiver undergoes multiple scattering and
propagates along various geometric paths associated with different
power losses, different time delays and direction of arrivals, as
pictured in Fig.\ref{rsf_points}. The spatial domain
$\calE\subseteq\IR^q$ accommodating the scattering objects
together with the signal transmitter and the receiver is called
the scattering environment, where $q\in\IN$ is the dimension of
the space under consideration, usually $q=2$ or $3$. In general,
an RSF, denoted by $S(t;\bfr,\bfv)$, may contain spatially
extensive scattering objects, which could be continuously
distributed in the real space and the velocity space, namely the
configuration space. Fig.\ref{rsf_points} only shows a special RSF
with discrete scattering objects, which may be the most adopted
model \cite{Sadowsky98,Fuhl98}. A discrete RSF is often modeled by
a compound point process \cite{Fishman76,Snyder91} in the
configuration space $\calE\times\IR^q$,
\begin{equation}
S(t;\bfr,\bfv)=\sum_{k=3}^{N(t)}S_k\delta(\bfr-\bfr_k)
\delta(\bfv-\bfv_k), \label{Strv}
\end{equation}
where $\bfr=(r_1,\cdots,r_q)'\in\calE$ and
$\bfv=(v_1,\cdots,v_q)'\in\IR^q$ denote\footnote{In this paper,
$A'$ denotes the transpose of $A$, for any vector or matrix $A$.}
the position and the velocity respectively, and hence
$\{(\bfr_k,\bfv_k)\}_{k=3}^{N(t)}$ are configuration coordinates
for the scattering objects, which are the realization points of a
space-time point process in $[0,\infty)\times\calE\times\IR^q$.
Here $N(t)$ is the total number of points occurred up to time $t$
anywhere in $\calE\times\IR^q$. The scattering objects are indexed
by the integers $k\ge 3$, the indices $k=1,2$ are reserved for the
transmitter and the receiver respectively. For each $k\ge 3$,
$S_k$ is a complex-valued random variable associated with the
$k$th realization point, which represents the scattering response
of the corresponding scattering object to the signal. Although the
discrete RSF model is chosen here to simplify the discussion, it
should be pointed out that, a more general RSF model might, and
should be used when necessary, for example, when describing a
scattering environment with continuously distributed scattering
objects. In fact, all mathematical formulae with $S(t;\bfr,\bfv)$
involved in the present paper are formulated in such general forms
that they hold for any $\calD'(\calE\times\IR^q)$-valued process
$S(t;\bfr,\bfv)$, where $\calD'(\calE\times\IR^q)$ denotes the
Hilbert space of the complex-valued, linear functionals of
functions in $\calE\times\IR^q$, which are usually called
distributions or generalized functions
\cite{Gelfand64,Schwartz66}.

The RSF imprints itself on the channel response between a wave
transmitter and a signal receiver, all of which lie in the
scattering environment $\calE$. In general, both the transmitter
and the receiver could be extensive in space, as being the case
when antenna arrays are employed. Therefore we use
$x(t;\bfr_1,\bfv_1)$ and $y(t;\bfr_2,\bfv_2)$ to denote the
spatially extensive signal amplitudes of the transmitter and the
receiver respectively, both of which are complex-valued and square
integrable in the Lebesgue sense, namely, $x,y\in
L^2([0,\infty)\times\calE\times\IR^q,\calC)$, where $\calC$ is the
set of complex numbers. Recalling the theory of wave scattering
which says that the otherwise freely propagating signal wave from
a source is perturbed by the scattering objects, and the actual
wave in space is the superposition of the free-running wave and
all of the scattered waves. The Huygens-Fresnel principle
\cite{Jackson75,Goodman96} further states that the scattered wave
from each scattering object, often called the second wave, is
again a free-running wave coming out of the object, with an
amplitude proportional to that of the incident wave, and the
proportional coefficient is called the scattering response of the
object. In general, it is rather complicated to calculate the
actual wave in an exact manner, which involves the solution of an
integral equation or a series expansion using the Feynman diagram
\cite{Ishim78,Rossum99}. The complication has to do with the
multiple scattering effects. However, when the scattering objects
are sparsely distributed in space, their effect on the signal wave
may be easily calculated using the perturbation theory. In
particular, the first-order perturbation theory approximates the
actual wave by superposing the free-running wave and the second
waves excited by it at the scattering objects. For an obvious
reason, the first-order perturbation is also called the
single-bounce approximation. When considering only the
contribution of the second waves, the single-bounce approximation
results in an input-output relation of the spatially extensive
signals $x(t;\bfr_1,\bfv_1)$ and $y(t;\bfr_2,\bfv_2)$ in their
low-pass equivalent form \cite{Proakis00},
\begin{equation}
y(t;\bfr_2,\bfv_2)=\sum_{k=3}^{N(t)}\int_{\calE\times\IR^q}
x(t-\tau_{1k}-\tau_{k2};\bfr_1,\bfv_1)l(c\tau_{1k})S_kl(c\tau_{k2})
\exp(i2\pi\gamma_{1k2}t)d\bfr_1d\bfv_1, \label{in-out-Sk}
\end{equation}
where $\tau_{1k}$ and $\tau_{k2}$ are the time delays associated
with the signal propagation from the transmitter to the $k$th
object and from the $k$th object to the receiver respectively,
which satisfy the following equations,
\begin{eqnarray}
\tau_{1k}=\frac{|\bfr_1-\bfr_k(t-\tau_{k2})|}{c},~\tau_{k2}=
\frac{|\bfr_k(t-\tau_{k2})-\bfr_2|}{c},~\forall~k\ge 3.
\label{EQtau}
\end{eqnarray}
Also in the above equations, $c$ is the speed of the wave,
$l(c\tau)$ is the propagation loss that depends only on the
propagation distance $c\tau$, and
\begin{equation}
\gamma_{1k2}\defi\frac{f_0[\bfv_1-\bfv_k(t-\tau_{k2})]\cdot
[\bfr_1-\bfr_k(t-\tau_{k2})]}{c|\bfr_1-\bfr_k(t-\tau_{k2})|}+
\frac{f_0[\bfv_k(t-\tau_{k2})-\bfv_2]\cdot[\bfr_k(t-\tau_{k2})
-\bfr_2]}{c|\bfr_k(t-\tau_{k2})-\bfr_2|}, \label{defgamma1k2}
\end{equation}
is the Doppler frequency shift experienced by the signal with the
central carrier frequency $f_0$. Note that the time delays
$\{\tau_{k2}\}_{k\ge 3}$ are implicitly defined in equation
(\ref{EQtau}), although an explicit formula may be derived when
the kinetic motions of the scattering objects are given
explicitly. For example, if the velocity difference
$\bfv_k(t)-\bfv_k(t-\tau_{k2})$ is neglected for all $k\ge 3$,
that is to ignore the velocity variation during the time of the
wave propagating from the scattering object to the receiver, then
the following equation holds,
\begin{equation}
c\tau_{k2}=|\bfr_k(t)-\bfv_k(t)\tau_{k2}-\bfr_2|,~\forall~k\ge 3,
\label{EQtau'}
\end{equation}
which is easily solved to give an explicit formula,
\begin{equation}
\tau_{k2}=\frac{\sqrt{[c^2-|\bfv_k(t)|^2]|\bfr_k(t)-\bfr_2|^2+
\{[\bfr_k(t)-\bfr_2]\cdot\bfv_k(t)\}^2}-[\bfr_k(t)-\bfr_2]\cdot
\bfv_k(t)}{c^2-|\bfv_k(t)|^2},~\forall~k\ge 3. \label{EQtau''}
\end{equation}
Then the time delays $\{\tau_{1k}\}_{k\ge 3}$ are explicitly
computed using equations (\ref{EQtau}) and (\ref{EQtau''}).
Similarly, equation (\ref{defgamma1k2}) could be simplified if all
$\{\bfv_k\}_{k\ge 3}$ are treated as constants during the interval
$[t-\tau_{k2},t]$,
\begin{equation}
\gamma_{1k2}\defi\frac{f_0[\bfv_1-\bfv_k(t)]\cdot[\bfr_1-\bfr_k(t)
+\bfv_k(t)\tau_{k2}]}{c|\bfr_1-\bfr_k(t)+\bfv_k(t)\tau_{k2}|}+
\frac{f_0[\bfv_k(t)-\bfv_2]\cdot[\bfr_k(t)-
\bfv_k(t)\tau_{k2}-\bfr_2]}{c|\bfr_k(t)-\bfv_k(t)\tau_{k2}
-\bfr_2|},~\forall~k\ge 3. \label{defgamma1k2'}
\end{equation}
Neglecting the object velocity variation during the time of wave
propagation is often an excellent approximation to make,
especially when dealing with electromagnetic waves, which travel
at the ultimate speed in nature, $c=3.0\times 10^8$ m/s. Even for
ten kilometers of propagation distance, the time delay is merely
$1/30$ millisecond, during which no macroscopic scattering object
can change its speed for much. Furthermore, the highest speed of
most macroscopic objects is no more than a millionth of the light
speed. Consequently, equation (\ref{EQtau''}) may be simplified,
\begin{equation}
\tau_{k2}=\frac{|\bfr_k(t)-\bfr_2|}{c}-\frac{[\bfr_k(t)-\bfr_2]
\cdot\bfv_k(t)}{c^2},~\forall~k\ge 3, \label{EQtau'''}
\end{equation}
by neglecting terms of the second or higher powers of
$\bfv_k(t)/c$, $\forall~k\ge 3$. Or even simpler, while still an
extremely good approximation for the propagation of
electromagnetic wave, the time delays may be calculated as,
\begin{eqnarray}
\tau_{1k}=\frac{|\bfr_1-\bfr_k(t)|}{c},~\tau_{k2}=
\frac{|\bfr_k(t)-\bfr_2|}{c},~\forall~k\ge 3. \label{EQtau4}
\end{eqnarray}

To write the input-output relation in a more compact form, we
define
\begin{equation}
\tau_2(\bfr,\bfv;\bfr_2)\defi\frac{\sqrt{(c^2-|\bfv|^2)
|\bfr-\bfr_2|^2+[(\bfr-\bfr_2)\cdot\bfv]^2}-
(\bfr-\bfr_2)\cdot\bfv}{c^2-|\bfv|^2},~\forall~k\ge 3,
\label{tau2def}
\end{equation}
\begin{equation}
\tau_1(\bfr_1;\bfr,\bfv;\bfr_2)\defi\frac{|\bfr_1-\bfr+
\bfv\tau_2(\bfr,\bfv;\bfr_2)|}{c},~\forall~k\ge 3, \label{tau1def}
\end{equation}
\begin{equation}
\gamma(\bfr_1,\bfv_1;\bfr,\bfv;\bfr_2,\bfv_2)\defi
\frac{f_0(\bfv_1-\bfv)\cdot[\bfr_1-\bfr+
\bfv\tau_2(\bfr,\bfv;\bfr_2)]}{c|\bfr_1-\bfr+
\bfv\tau_2(\bfr,\bfv;\bfr_2)|}+\frac{f_0(\bfv-\bfv_2)\cdot
[\bfr-\bfv\tau_2(\bfr,\bfv;\bfr_2)-\bfr_2]}{c|\bfr-
\bfv\tau_2(\bfr,\bfv;\bfr_2)-\bfr_2|}, \label{gamdef}
\end{equation}
\begin{equation}
L(\bfr_1,\bfv_1;\bfr,\bfv;\bfr_2,\bfv_2)\defi
l[c\tau_1(\bfr_1;\bfr,\bfv;\bfr_2)]l[c\tau_2(\bfr,\bfv;\bfr_2)]
\exp[i2\pi\gamma(\bfr_1,\bfv_1;\bfr,\bfv;\bfr_2,\bfv_2)],
\label{Ldef}
\end{equation}
then we use equation (\ref{Strv}), together with equations
(\ref{EQtau''}), and (\ref{defgamma1k2'}) to rewrite equation
(\ref{in-out-Sk}) as,
\begin{eqnarray}
y(t;\bfr_2,\bfv_2)&=&\int_{(\calE\times\IR^q)^2}
x[t-\tau_1(\bfr_1;\bfr,\bfv;\bfr_2)-
\tau_2(\bfr,\bfv;\bfr_2);\bfr_1,\bfv_1]\nonumber\\
&\times&L(\bfr_1,\bfv_1;\bfr,\bfv;\bfr_2,\bfv_2)
S[t-\tau_2(\bfr,\bfv;\bfr_2);\bfr,\bfv]
d\bfr_1d\bfv_1d\bfr d\bfv. \label{in-out}
\end{eqnarray}
This equation shall serve as the fundamental input-output relation
describing the propagation of spatially extensive signals in an
RSF, up to the single-bounce approximation. It is evident that the
received signal $y(t;\bfr_2,\bfv_2)$ is a bilinear functional of
$x(t;\bfr_1,\bfv_1)$ and $S(t;\bfr,\bfv)$, for any fixed
$(t;\bfr_2,\bfv_2)$. More specifically, $y(t;\bfr_2,\bfv_2)$ is a
linear functional of $x(t;\bfr_1,\bfv_1)$, if
$L(\bfr_1,\bfv_1;\bfr,\bfv;\bfr_2,\bfv_2)S(t;\bfr,\bfv)$ is
regarded as an integration kernel; or $y(t;\bfr_2,\bfv_2)$ could
be viewed as a linear functional of $S(t;\bfr,\bfv)$, with
$x(t;\bfr_1,\bfv_1)L(\bfr_1,\bfv_1;\bfr,\bfv;\bfr_2,\bfv_2)$
serving as an integration kernel.

\section{The Dynamics of RSF}
Let us consider the case in which the set of scattering objects is
fixed in time and the scattering response of each object is
time-independent. Nevertheless, the RSF is still time-varying when
the objects are subject to kinetic motion. We shall employ an RSF
model so general that the scattering objects are subject to
accelerations that depend upon the position. Specifically, the
motion of the $k$th scattering object, $k\ge 3$, is described by a
differential equation,
\begin{eqnarray}
d\bfr_k&=&\bfv_kdt, \label{drEQ}\\
d\bfv_k&=&f(\bfr_k)dt, \label{dvEQ}
\end{eqnarray}
where $f(\bfr)$ is an $\IR^q$-valued function representing the
deterministic acceleration exerted upon the scattering objects.
Since the total number of scattering objects within the
environment $\calE$ is assumed to be fixed in time, it is a random
variable $N=N(\omega)$. As mentioned before, all random variables
and processes are defined on the ``grand'' probability space
$(\Omega,\calF,P)$. Let
\begin{equation}
S(t;\bfr,\bfv)\defi\sum_{k=3}^NS_k\delta(\bfr-\bfr_k)
\delta(\bfv-\bfv_k), \label{Strvab}
\end{equation}
then $S(t;\bfr,\bfv)$ is a space-time compound point process
\cite{Fishman76,Snyder91} in the space
$[0,\infty)\times\calE\times\IR^q$. Let $U$ denotes the Hilbert
space of all generalized functions in $\calE\times\IR^q$, namely
$U\defi\calD'(\calE\times\IR^q)$, then $S(t;\bfr,\bfv)$ may also
be viewed as a $U$-valued random process indexed by
$t\in[0,\infty)$. Next we attempt to derive a partial differential
equation for $S(t;\bfr,\bfv)$ using equations (\ref{drEQ}),
(\ref{dvEQ}) and (\ref{Strvab}).

While explicitly parameterized by $\bfr$ and $\bfv$, the RSF
$S(t;\bfr,\bfv)$ of equation (\ref{Strvab}) is implicitly a
function of the processes $\{(\bfr_k,\bfv_k)\}_{k=3}^N$, described
by the differential equations (\ref{drEQ}) and (\ref{dvEQ}). By
differentiating equation (\ref{Strvab}), it is obtained that,
\begin{eqnarray}
dS(t;\bfr,\bfv)&=&\sum_{k}S_k\delta(\bfv-\bfv_k)d\delta(\bfr-
\bfr_k)+\sum_{k}S_k\delta(\bfr-\bfr_k)d\delta(\bfv-\bfv_k)
\nonumber\\
&=&\sum_{k}S_k\delta(\bfv-\bfv_k)\frac{\partial\delta(\bfr-
\bfr_k)}{\partial\bfr_k}\bfv_kdt+\sum_{k}S_k\delta(\bfr-\bfr_k)
\frac{\partial\delta(\bfv-\bfv_k)}{\partial\bfv_k}f(\bfr_k)dt
\nonumber\\
&=&-\sum_{k}S_k\delta(\bfv-\bfv_k)\frac{\partial\delta(\bfr-
\bfr_k)}{\partial\bfr}\bfv_kdt-\sum_{k}S_k\delta(\bfr-\bfr_k)
\frac{\partial\delta(\bfv-\bfv_k)}{\partial\bfv}f(\bfr_k)dt
\nonumber\\
&=&-\sum_{k}S_k\delta(\bfv-\bfv_k)\frac{\partial\delta
(\bfr-\bfr_k)}{\partial\bfr}\bfv dt-\sum_{k}S_k\delta(\bfr-\bfr_k)
\frac{\partial\delta(\bfv-\bfv_k)}{\partial\bfv}f(\bfr)dt,
\label{1stdS}
\end{eqnarray}
for all $k\ge 3$, where in the last step, two identities have been
used,
\begin{eqnarray}
\delta(\bfv-\bfv_k)\bfv_k&\equiv&\delta(\bfv-\bfv_k)\bfv,
~\forall~k\ge 3,\\
\delta(\bfr-\bfr_k)f(\bfr_k)&\equiv&\delta(\bfr-\bfr_k)f(\bfr),
~\forall~k\ge 3.
\end{eqnarray}
The partial differential operator $\partial/\partial\bfr$ is
defined as a row vector,
\begin{equation}
\frac{\partial}{\partial\bfr}\defi\left[\frac{\partial}{\partial
r_1},\cdots,\frac{\partial}{\partial r_q}\right],
\end{equation}
and the operator $\partial/\partial\bfv$ is similarly defined.
Note that $f(\bfr)$ is valued as a $q$-dimensional column vector.
When a row vector is followed by a column vector, it is understood
that the scalar product in the space $\IR^q$ takes place. Noticing
two more identities,
\begin{eqnarray}
\frac{\partial
S(t;\bfr,\bfv)}{\partial\bfr}&=&\sum_kS_k\delta(\bfv-\bfv_k)
\frac{\partial\delta(\bfr-\bfr_k)}{\partial\bfr},\\
\frac{\partial
S(t;\bfr,\bfv)}{\partial\bfv}&=&\sum_kS_k\delta(\bfr-\bfr_k)
\frac{\partial\delta(\bfv-\bfv_k)}{\partial\bfv},
\end{eqnarray}
one may easily simplify equation (\ref{1stdS}) into a partial
differential equation governing the RSF,
\begin{equation}
dS(t;\bfr,\bfv)=-\frac{\partial S(t;\bfr,\bfv)}{\partial\bfr}\bfv
dt-\frac{\partial S(t;\bfr,\bfv)}{\partial\bfv}f(\bfr)dt.
\label{2nddS}
\end{equation}
Define a linear operator $A:U\rightarrow U$, such that for any
$\xi\in U$,
\begin{equation}
A\xi(\bfr,\bfv)\defi-\frac{\partial\xi}{\partial\bfr}\bfv-
\frac{\partial\xi}{\partial\bfv}f(\bfr),
\label{Adef}
\end{equation}
then equation (\ref{2nddS}), together with a proper initial
condition, is immediately recognized as an infinite dimensional
linear system \cite{Curtain78,Sawaragi78} in the Hilbert space
$U$,
\begin{eqnarray}
dS(t)&=&AS(t)dt, \label{dSfinal1}\\
S(0)&=&S_0, \label{dSfinal2}
\end{eqnarray}
where $S_0$ is an $(\calF,\calB(U))$-measurable random variable,
$\calB(U)$ is a Borel $\sigma$-algebra of subsets of $U$. It is to
be stressed again that the dynamic equations (\ref{dSfinal1}) and
(\ref{dSfinal2}) are applicable to any RSF $S(t)\in
U=\calD'(\calE\times\IR^q)$, although the derivation of the
equations is exemplified by the special case of discrete RSF for
simplicity.

\section{Signal Detection}
As discussed in section \ref{waveinrsf}, when a spatially
extensive signal $x(t;\bfr_1,\bfv_1)$ is transmitted, a spatially
extensive receiver will catch a signal $y(t;\bfr_2,\bfv_2)$, which
may be degraded by a white Gaussian noise. Let
\begin{equation}
Y(t;\bfr_2,\bfv_2)\defi\int_0^ty(s;\bfr_2,\bfv_2)ds+Z(t;\bfr_2,\bfv_2),
\end{equation}
where $Z(t;\bfr_2,\bfv_2)$ is an $H$-valued Wiener process
representing the observation noise, with $H\defi
L^2(\calE\times\IR^q)$, that is, the space of square Lebesgue
integrable functions in $\calE\times\IR^q$. It is also assumed
that the covariance operator of $Z(t)$, denoted by $R$, is
uniformly positive definite. Using equation (\ref{in-out}), with
the simplified calculation of delay times in (\ref{EQtau4}), we
can write,
\begin{eqnarray}
dY(t;\bfr_2,\bfv_2)&=&\left[\int_{(\calE\times\IR^q)^2}x\left(t-
\frac{|\bfr_1-\bfr|+|\bfr-\bfr_2|}{c};\bfr_1,\bfv_1\right)
L(\bfr_1,\bfv_1;\bfr,\bfv;\bfr_2,\bfv_2)\right.\nonumber\\
&\times&\left.S\left(t-\frac{|\bfr-\bfr_2|}{c};\bfr,\bfv\right)
d\bfr_1d\bfv_1d\bfr d\bfv\right]dt+dZ(t;\bfr_2,\bfv_2),
\label{observation1'}\\
Y(0)&=&0, \label{observation2'}
\end{eqnarray}
which is the signal detection equation, or called the observation
equation. With the observation data $\{Y(s)\}_{0\le s<t}$, two
estimation problems emerge, depending upon the actual application
goal:
\begin{description}
\item[1)] estimating and predicting $S(t)$, when $\{x(s)\}_{0\le
s\le t}$ is known; \item[2)] estimating $x(t)$, when
$\{S(t)\}_{0\le s\le t}$ is known.
\end{description}
The first problem is related to various applications of inverse
scattering, such as radar object detection, channel estimation and
prediction for wireless communications; while the second problem
is specific to signal detection, especially to the optimal design
of a receiver in communications. In the present paper, we are
interested in the first estimation problem, in which the signal
$x(t;\bfr_1,\bfv_1)$ is regarded as a known quantity. Define a
time-dependent linear operator $G(t):U\rightarrow H$, such that
for any time-varying function $\xi(t)\in U$, $\forall~t\ge 0$,
\begin{eqnarray}
&&G(t)\xi(t)(\bfr_2,\bfv_2)\defi\nonumber\\
&&\int_{(\calE\times\IR^q)^2}x\left(t-\frac{|\bfr_1-\bfr|+
|\bfr-\bfr_2|}{c};\bfr_1,\bfv_1\right)L(\bfr_1,\bfv_1;
\bfr,\bfv;\bfr_2,\bfv_2)\xi\left(t-\frac{|\bfr-\bfr_2|}{c};
\bfr,\bfv\right)d\bfr_1d\bfv_1d\bfr d\bfv, \label{Gdef}
\end{eqnarray}
then the observation equations (\ref{observation1'}) and
(\ref{observation2'}) are of the form,
\begin{eqnarray}
dY(t)&=&G(t)S(t)dt+dZ(t),\label{observation1}\\
Y(0)&=&0,\label{observation2}
\end{eqnarray}
where $Y(t)$ and $Z(t)$ are $H$-valued random processes, with
$H=L^2(\calE\times\IR^q)$, while $G(t)$ is a time-dependent linear
operator in $\calL_{HS}(U,H)$, with the time-dependence originated
from the known and time-varying signal $x(t)$, $\calL_{HS}(U,H)$
being the space of all Hilbert-Schmidt operators from $U$ to $H$.

The system equations (\ref{dSfinal1}), (\ref{dSfinal2}) and the
observation equations (\ref{observation1}), (\ref{observation2})
constitute a nonlinear filtering problem, that is to compute the
conditional probability,
\begin{equation}
\Pi(t)(\psi(t))\defi\E[\psi(t,S(t))|Y(s),0\le s\le t],
\end{equation}
where $\psi(t,\xi)\in C_b^{1,2}([0,\infty)\times U)$ is a given
test function, which is bounded and has continuous first and
second order Fr\'{e}chet derivatives with respect to $t$ and $\xi$
respectively. The essence of the nonlinear filtering problem is to
obtain the optimal estimation of the interested quantity
$\psi(t,S(t))$, based on the observation data up to time $t$,
$\{Y(s)\}_{0\le s\le t}$. In most practical applications, it is
sufficient to obtain a dynamic evolution equation for the
estimated $\psi(t,S(t))$, usually in the form a differential
equation driven by the observation data, which can be implemented,
or simulated, by a state-possessing machine or circuit.
Accordingly, a dynamic evolution equation for the estimated
quantity may be regarded as a solution to the nonlinear filtering
problem. Thanks to the greatly advanced theories of nonlinear
filtering and of finite and infinite dimensional systems, such
dynamic evolution equations have been well established as
solutions to the nonlinear filtering problems.

\section{Channel Estimation and Prediction}
For the general theory of nonlinear filtering, the readers are
referred to the excellent textbooks
\cite{Krishnan84,Bensoussan92,Ahmed98} and references therein,
especially the original papers. Here we shall formulate the
problem in a less general form, but to suit our applications
better. We are interested in an inverse scattering problem, which
is recast as, given,
\begin{eqnarray}
&&dS(t)=AS(t)dt,~~~~~~~~~~~~~~~~S(0)=S_0,\\ \label{system}
&&dY(t)=G(t)S(t)dt+dZ(t),~~Y(0)=0, \label{observation}
\end{eqnarray}
to compute,
\begin{equation}
\Pi(t)(\psi(t))=\E[\psi(t,S(t))|Y(s),0\le s\le t], \label{target}
\end{equation}
where $Z(t)$ is an $H$-valued Wiener process, with covariance
operator $R$, which is uniformly positive definite. $S_0$ is an
$(\calF,\calB(U))$-measurable random variable with probability
measure $\Pi_0$. $Z(t)$ and $S_0$ are independent. The operators
$A$ and $G(t)$ are defined in (\ref{Adef}) and (\ref{Gdef})
respectively. Although we have successfully formulated the general
inverse scattering problem as a nonlinear filtering problem for an
infinite dimensional linear system, we shall not present the
general solution to the problem, in order to limit the scope and
length of the discussion. The general solution involves much
elaborate notations and concepts in functional analysis, as well
as somewhat tedious arguments of convergence, which does not seem
to fit the purpose of the present paper. Interested readers,
however, are referred to the literature \cite{Prato92,Ahmed97}.

What we shall focus here, is a special case of discrete RSF with
the accelerations of the scattering objects being neglected,
namely, all objects are moving with constant velocities. The goal
of inverse scattering is to find out the initial position $\bfr_k$
and velocity $\bfv_k$, as well as the attenuation coefficient
$S_k$ to the wave, for each scattering object $k$, $3\le k\le N$,
based upon the received signal $\{Y(s)\}_{0\le s\le t}$ and the
known transmitted signal $\{x(s)\}_{0\le s\le t}$. However, the
total number of scattering objects $N$ is usually random, and to
pin down the configuration coordinates
$\{(\bfr_k,\bfv_k)\}_{k=2}^N$ from scratch is highly difficult. A
more practical approach is to discretize the initial configuration
space into small regions and label each of them by an integer $n$,
$1\le n\le M$, then to assign a complex-valued random variable
$\sfS_n$ as the scattering response to the center of the $n$th
region, which is coordinated by $(\sfr_n,\sfv_n)$. The idea is,
when the initial configuration space is divided into sufficiently
small regions, the scattering objects within any region $n$, $1\le
n\le M$, become non-resolvable, their scattering responses
superposed together is represented by a single scattering point
sitting at the center of the region with a scattering response
$\sfS_n$. The result of the discretization is to remove the
coordinates of the scattering objects in the initial configuration
space from the set of the unknowns, but to let the complex random
variables $\{\sfS_n\}_{n=1}^M$ bear all the information about the
scattering environment. The probability law of
$\{\sfS_n\}_{n=1}^M$ may be obtained from some experimental data,
or calculated from an empirical model of the distribution of
scattering objects in an actual geographical environment
\cite{Sadowsky98,Fuhl98}. For instance, if the distribution of
objects is quite sparce in an area, then the probability of
$\sfS_n=0$ is high, for any $1\le n\le M$. Let
\begin{equation}
\sfS\defi(\sfS_1,\sfS_2,\cdots,\sfS_M)',
\end{equation}
then $\sfS$ is a random variable in $\calC^M$, whose probability
law is known and denoted by $\Pi_0$ just as before. Since the
scattering response of each object is assumed time-invariant, the
RSF $S(t)$, now $\calC^M$-valued, is time-independent,
\begin{equation}
dS(t)=0,~~S(0)=\sfS. \label{dSis0}
\end{equation}
The transmitter and receiver are also discrete in the
configuration space, which consist $I$ and $J$ antenna elements
located at $\{(\sfr_{1i},\sfv_{1i})\}_{i=1}^I$ and
$\{(\sfr_{2j},\sfv_{2j})\}_{j=1}^J$ respectively. As a
consequence, the transmitted signal $x(t)$ is a $\calC^I$-valued
signal, while the Wiener noise $Z(t)$ at the receiver and the
received signal $Y(t)$ are $\calC^J$-valued random processes,
\begin{eqnarray}
x(t)&\defi&(x_1(t),x_2(t),\cdots,x_I(t))',\\
Y(t)&\defi&(Y_1(t),Y_2(t),\cdots,Y_J(t))',\\
Z(t)&\defi&(Z_1(t),Z_2(t),\cdots,Z_J(t))'.
\end{eqnarray}
Again let $R(t)$ denote the covariance matrix of the Wiener
process $Z(t)$, which is uniformly positive definite. Define a
$J\times M$ matrix $\sfG(t)=\{g_{jn}(t)\}_{1\le j\le J,1\le n\le
M}$, such that,
\begin{equation}
g_{jn}(t)\defi\sum_{i=1}^Ix\left(t-\frac{|\sfr_{1i}-\sfr_n|+
|\sfr_n-\sfr_{2j}|}{c};\sfr_{1i},\sfv_{1i}\right)
L(\sfr_{1i},\sfv_{1i};\sfr_n,\sfv_n;\sfr_{2j},\sfv_{2j}),
~\forall~1\le j\le J,~\forall~1\le n\le M,
\end{equation}
where the function $L$ is defined in (\ref{Ldef}), then the
discrete version of the observation equation (\ref{observation1'})
is formulated as,
\begin{equation}
dY(t)=\sfG(t)\sfS dt+dZ(t),~~Y(0)=0. \label{dis_obs}
\end{equation}

With the trivial system equation (\ref{dSis0}) and the simple
observation equation (\ref{dis_obs}), the inverse scattering
problem is to compute,
\begin{equation}
\Pi(t)(\psi(t))=\E[\psi(t,S(t))|Y(s),0\le s\le
t]=\E[\psi(t,\sfS)|Y(s),0\le s\le t],
\end{equation}
where $\psi(t,\xi)\in C_b^{1,2}([0,\infty)\times\calC^M,\calC)$ is
a given test function, which is bounded and has continuous first
and second order derivatives with respect to $t$ and $\xi$
respectively. For the convenience in presenting the following
filtering equations, let
$B:[0,\infty)\times\calC^M\rightarrow\calC^J$ be a function such
that,
\begin{eqnarray}
B(t,\xi)&\defi&\sfG(t)\xi,~\forall~t\ge 0,~\forall~\xi\in\calC^M,\\
B'(t,\xi)&\defi&\xi'\sfG'(t),~\forall~t\ge
0,~\forall~\xi\in\calC^M.
\end{eqnarray}
It is a trivial matter to verify the conditions in order to apply
the Kushner-Stratonovitch equation \cite{Krishnan84,Bensoussan92},
\begin{eqnarray}
d\Pi(t)(\psi(t))&=&\Pi(t)\left(\frac{\partial\psi}{\partial
t}\right)dt+\left[\Pi(t)(B'(t)\psi(t))-\Pi(t)(\psi(t))\Pi(t)(B'(t))
\right]\nonumber\\
&\times&R^{-1}(t)[dY(t)-\Pi(t)(B(t))dt],~\mbox{\rm almost surely},
~\forall~t>0,\\
\Pi(0)(\psi(0))&=&\Pi_0(\psi(0))=\int_{\calC^M}
\psi(0,\xi)d\Pi_0(\xi).
\end{eqnarray}
Another well known solution to the nonlinear filtering problem is
the Zakai equation \cite{Krishnan84,Bensoussan92}, which governs
the so-called unnormalized conditional probability
$p(t)(\psi(t))$,
\begin{eqnarray}
dp(t)(\psi(t))&=&p(t)\left(\frac{\partial\psi}{\partial
t}\right)dt+p(t)(B'(t)\psi(t))
R^{-1}(t)dY(t),~\mbox{\rm almost surely},~\forall~t>0,\\
p(0)(\psi(0))&=&\Pi_0(\psi(0))=\int_{\calC^M}\psi(0,\xi)d\Pi_0(\xi).
\end{eqnarray}
The relation between $p(t)$ and $\Pi(t)$ is given by,
\begin{equation}
\Pi(t)(\psi)=\frac{p(t)(\psi)}{p(t)(1)},~\forall~\mbox{\rm test
function}~\psi,
\end{equation}
where $p(t)(1)$ satisfies,
\begin{eqnarray}
dp(t)(1)&=&p(t)(B'(t))R^{-1}(t)dY(t),~\mbox{\rm almost surely},
~\forall~t>0,\\
p(0)(1)&=&1.
\end{eqnarray}

As a simpler example, consider a flat-fading channel from a single
transmitter element to a single receiver element, where the
frequency bandwidth of the channel is sufficiently narrow that the
time delays among the signal paths are insignificant, however, the
channel response experiences a fast oscillation due to the motion
of the receiver, or the transmitter, or the scattering objects.
The goal is to estimate the channel state and to predict the
channel response, based on the known transmitted signal and the
data of received signal. This problem has been tackled before with
the help of various spectral estimation techniques
\cite{Eyceoz98,Duel-Hallen00,Hwang98}, which do not guarantee the
optimality of the solutions. It is therefore of great interest to
formulate the problem using the language of nonlinear filtering,
and to see what kind of optimal solution may be obtained. The
observation equation, namely, the channel input-output relation,
has a very simple low-pass equivalent form \cite{Proakis00},
\begin{equation}
dY(t)=x(t)h(t)dt+dZ(t),~~Y(0)=0,
\end{equation}
where $x(t)$ and $Y(t)$ are the transmitted signal and the
integrated version of the received signal respectively, and $h(t)$
is the time-varying channel response, all $\calC$-valued, while
$Z(t)$ is a $\calC$-valued Wiener process with a uniformly
positive definite covariance function $R(t)$. The channel response
$h(t)$ is a superposition of many Doppler components,
\begin{equation}
h(t)=\sum_{k=1}^Ka_k\exp(i2\pi f_kt),~~-f_{\max}\le
f_1<f_2<\cdots<f_K\le f_{\max},
\end{equation}
where $f_{\max}$ is the maximum Doppler frequency shift, and
$\{f_k\}_{k=1}^K$ constitutes a partition of the frequency band
$[-f_{\max},f_{\max}]$, and $a_k\in\calC$ is the constant
amplitude of the $k$th Doppler component, $1\le k\le K$. It is
assumed that the partition $\{f_k\}_{k=1}^K$ is sufficiently fine
to justify the discrete approximation to the Doppler shifts.
Define a function $F:[0,\infty)\times\calC^K\rightarrow\calC$,
such that,
\begin{equation}
F(t,\xi)\defi x(t)\sum_{k=1}^K\exp(i2\pi
f_kt)\xi_k,~\forall~\xi=(\xi_1,\xi_2,\cdots,\xi_K)'\in\calC^K.
\end{equation}
The nonlinear filtering problem for channel estimation and
prediction is formulated as, given,
\begin{eqnarray}
&&\sfa\defi(a_1,a_2,\cdots,a_K)'\in\calC^K,~\mbox{\rm with
probability law}~\Pi_0,\\
&&dY(t)=F(t,\sfa)dt+dZ(t),~~Y(0)=0,
\end{eqnarray}
to compute,
\begin{equation}
\Pi(t)(\psi(t))=\E[\psi(t,\sfa)|Y(s),0\le s\le t],
\end{equation}
where $\psi\in C^{1,2}_b([0,\infty)\times\calC^K,\calC)$ is a test
function. Again, the solution is given by the
Kushner-Stratonovitch equation,
\begin{eqnarray}
d\Pi(t)(\psi(t))&=&\Pi(t)\left(\frac{\partial\psi}{\partial t}
\right)dt+\left[\Pi(t)(F(t)\psi(t))-\Pi(t)(\psi(t))\Pi(t)(F(t))
\right]\nonumber\\
&\times&R^{-1}(t)[dY(t)-\Pi(t)(F(t))dt],~\mbox{\rm almost
surely},~\forall~t>0,\\
\Pi(0)(\psi(0))&=&\Pi_0(\psi(0))=\int_{\calC^K}\psi(0,\xi)
d\Pi_0(\xi),
\end{eqnarray}
or by the Zakai equation,
\begin{eqnarray}
dp(t)(\psi(t))&=&p(t)\left(\frac{\partial\psi}{\partial
t}\right)dt+p(t)(F(t)\psi(t))
R^{-1}(t)dY(t),~\mbox{\rm almost surely},~\forall~t>0,\\
p(0)(\psi(0))&=&\Pi_0(\psi(0))=\int_{\calC^K}\psi(0,\xi)
d\Pi_0(\xi),
\end{eqnarray}
then the probability $\Pi(t)(\psi(t))$ is calculated as,
\begin{equation}
\Pi(t)(\psi(t))=\frac{p(t)(\psi(t))}{p(t)(1)}.
\end{equation}

\section{Conclusion}
It has been argued that the input-output response of a wireless
channel in a given scattering environment is actually
deterministic in nature, conditioned on the positions and
velocities of the scattering objects, as well as the locations and
velocities of the transmitter and receiver in the environment.
Using a physical model of wave propagation, the channel response
is calculated explicitly as a functional of the random scattering
field. Consequently, the best choice of the channel state
information for wireless channels may be the state of the random
scattering field, which can remain unchanged for a sufficiently
long time, even for the ``fast-fading channels'' in the
conventional sense. As far as the problem of channel estimation
and prediction is concerned, there seems to be no fundamental
limit imposed by the physics of wave scattering, no matter how
short is the carrier wavelength and how fast the mobiles move as
long as the inter-object separation is much larger than the
distance that any mobile can travel within the time duration of
interest. Only practical limits may rise from the speed and
complexity of the actual signal processing circuits. The problems
of inverse scattering, formulated as nonlinear filtering problems,
are solved in terms of dynamical evolution equations for the
estimated quantities. However, the filtering equations cannot be
implemented with little difficulty using today's signal processing
circuits. Both better numerical algorithms and more advanced
electronics are desired, in order to take fully the advantage of
the nonlinear filters in practical applications.

\end{document}